\documentclass[20pt, letterpaper, hyper]{JHEP3}
\pdfoutput=1

\let\ifpdf\relax

\usepackage{ifpdf}
\usepackage[utf8]{inputenc}
\usepackage{epic,eepic}
\usepackage{amsmath}
\usepackage{amssymb,amsfonts}
\usepackage{graphicx}

\title{On spontaneous breaking of conformal symmetry by probe flavour D-branes}

\author{Omer Ben-Ami${}^1$, Stanislav Kuperstein${}^2$, Jacob Sonnenschein${}^1$,
     \\
     ${}^1$ \textit{Raymond and Beverly Sackler Faculty of Exact Sciences \\
         School of Physics and Astronomy \\
         Tel-Aviv University, Ramat-Aviv 69978, Israel}
    \\
    ${}^2$ \textit{Institut de Physique Th\'{e}orique, CEA Saclay, CNRS URA 2306 \\F-91191 Gif-sur-Yvette, France}\\

\texttt{e-mails}: \textsf{omerben@post.tau.ac.il, stanislav.kuperstein@cea.fr, cobi@post.tau.ac.il}

}

\abstract{We explore the possibilities of breaking  conformal symmetry spontaneously by introducing flavour branes into conformal holographic backgrounds in the probe limit. A prototype model  of such a mechanism is based on placing $D7$-$\bar{D}7$ configuration in the Klebanov-Witten conifold based  model. In this paper we generalize this model. We conjecture on the required topology of the backgrounds and the corresponding probe brane embeddings. We identify several models that obey these requirements and admit spontaneous breaking of conformal invariance.
These include type IIB conifold based examples, dual to defect field theories based on the conifold, and type IIA constructions based on the ABJM model.
We identify the dilaton, the corresponding Goldstone boson, discuss its effective action and address the ``$a$-term". We briefly discuss the relevance of these models to the pseudo dilaton.

}

\preprint{TAUP-2978/13}

\begin{document}

%
%
%
%
%
%
%
%
%
%
%
%
%
%

\tableofcontents
\setcounter{footnote}{0}

\section{Introduction}
\label{sec:Intro}

Known examples of spontaneous conformal symmetry breaking are scarce. Although classically it may not be hard to find conformal invariant (interacting) field theories, it becomes a highly non-trivial task at the quantum level. This is due to the fact that one has to introduce a scale in order to regularize the theory. This scale can explicitly break the symmetry by telling us that marginal operators are not exactly marginal ({\it i.e.} QCD's $\beta$-function) or, rather miraculously, tell us that they are (the perturbative Banks-Zaks \cite{Banks:1981nn} interacting fixed point).

Finding such conformal theories is the first step. The second is to break the symmetry spontaneously. This means that the deformation of the interacting conformal field theory, initiating an RG-flow, should be a VEV deformation. VEV deformations with flat directions are especially hard to come by in conformal theories where ``Mexican hat" models are, of course, out of the question. They do come by, however, in supersymmetric conformal theories where flat directions are present also at the quantum level. The prime example being $\mathcal{N}=4$ SUSY with an RG flow from $SU(N)$ to $SU(N-1)$, initiated by giving a VEV to one of the scalars. Integrating out the massive fields, the theory flows from one fixed point to the other.
Although not necessarily perturbative, in these models only operators that already appear in the Lagrangian acquire a VEV.
There is another well known possibility of breaking a symmetry spontaneously - by strong coupling effects. This is usually also referred to as dynamical symmetry breaking. This is what happens in nature in chiral symmetry breaking where the quark condensate is the order parameter for the symmetry breaking. Strongly coupled conformal theories can in principle display the same behavior, thus breaking conformal symmetry spontaneously.

Generally speaking, there is no reason why there should be a flat potential for the scalar(s) in the theory (whether they are perturbative and already appear in the UV Lagrangian or a mesonic like operator). Supersymmetry, however, provides a mechanism which preservers the flat directions and allows for the scalar to receive an arbitrary VEV thus breaking the symmetry spontaneously. The models we explore break supersymmetry but still allow for the conformal symmetry to be broken spontaneously. We suspect that this is an artifact of our probe approximation and all corrections to the potential should appear once backreaction is included.

Our main motivation is to search for holographic models that admit spontaneous breaking of conformal invariance in a certain type of holographic models.

In general, spontaneous breaking of conformal symmetry in a holographic gravity setup can be done either by analysing holographic gravitational background with (or without) additional bulk fields or by embedding $N_\textrm{f}$ flavour D-branes in such backgrounds. Examples of the former approach can be found in \cite{Hoyos:2012xc},\cite{Hoyos:2013gma} where domain wall geometries in arbitrary number of dimensions interpolating between two AdS spaces, were analysed.
In this paper we follow solely the latter, D-brane embedding approach. Moreover, we will consider only the probe limit, where one can ignore the backreaction of the embedded D-branes on the geometry. This is easily achieved by taking the $N_\textrm{f} \ll N_\textrm{c}$ limit and in our discussion we will restrict ourself only to the $N_\textrm{f}=1$ case. We use the terminology "flavour" branes for branes which reach the UV boundary (unlike gauge branes) and where in the low-energy limit the interaction between flavour and gauge(color) strings vanishes. The VEV deformations we are interested in will then correspond to quark anti-quark condensates. This type of spontaneous conformal symmetry breaking remains rather unexplored in holography.

\begin{figure}[h]
\centering
\includegraphics[totalheight=0.2\textheight]{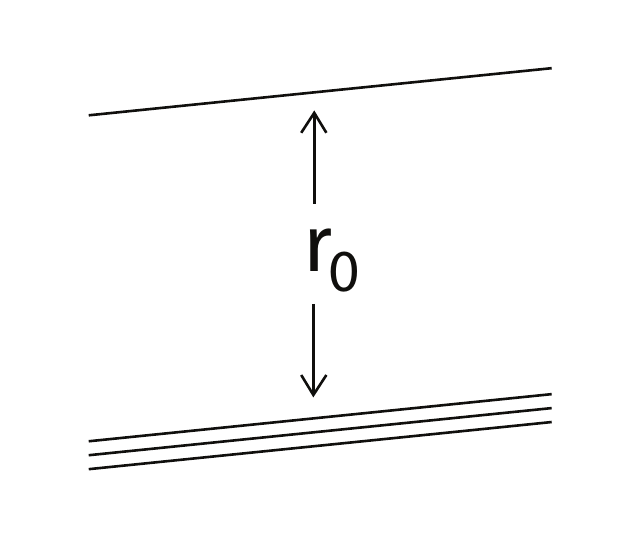}
\caption{A brane is separated from the background stack of $D3$ branes before going to the near horizon limit. Dual to the field theory Coulomb branch}
\label{fig:DpSeparation}
\end{figure}

We have already mentioned the RG flow from $SU(N)$ to $SU(N-1)$. The $\mathcal{N}=4$ $SU(N)$ CFT is holographically dual to the $AdS_5\times S^5$ geometry, which arises in the near horizon limit of a stack of $N$ backreacting D3-branes. The moduli space consists of giving a VEV to one of the three complex scalars. Taking a new near horizon limit the massive modes decouple again and so we are left with the gravity dual of the $SU(N-1)$ theory. The radial fluctuation of the separated brane will correspond to the Goldstone boson of the theory. This and other similar models will be briefly mentioned below when discussing the dilaton action. This configuration is depicted in Figure \ref{fig:DpSeparation}. This type of embeddings probe the geometry with gauge branes and the purpose of this paper is to explore flavour branes embeddings.

The prime example of a probe flavour brane model with spontaneously broken conformal symmetry was introduced in \cite{Kuperstein:2008cq}. In this setup D7 branes are embedded in the  Klebanov-Witten (KW) background \cite{Klebanov:1998hh},  $AdS_5 \times T^{1,1}$. The background is dual to an $\mathcal{N}=1$ SCFT with an $SU(N) \times SU(N)$ gauge group. The AdS geometry arises from the near horizon limit of D3 branes sitting at the (singular) apex of the $6d$ conifold. $T^{1,1}$ is the $5d$  conifold's base and has $S^3 \times S^2$ topology. The D7 brane analysed in \cite{Kuperstein:2008cq} wraps $AdS_5$ and the 3-sphere.
As a result, it looks like a point on the 2-sphere and in order to guarantee the tadpole cancellation one has to add an anti D7 annihilating the total D7 charge on the $S^2$.

\begin{figure}
	\centering
        \includegraphics[width=0.5\textwidth]{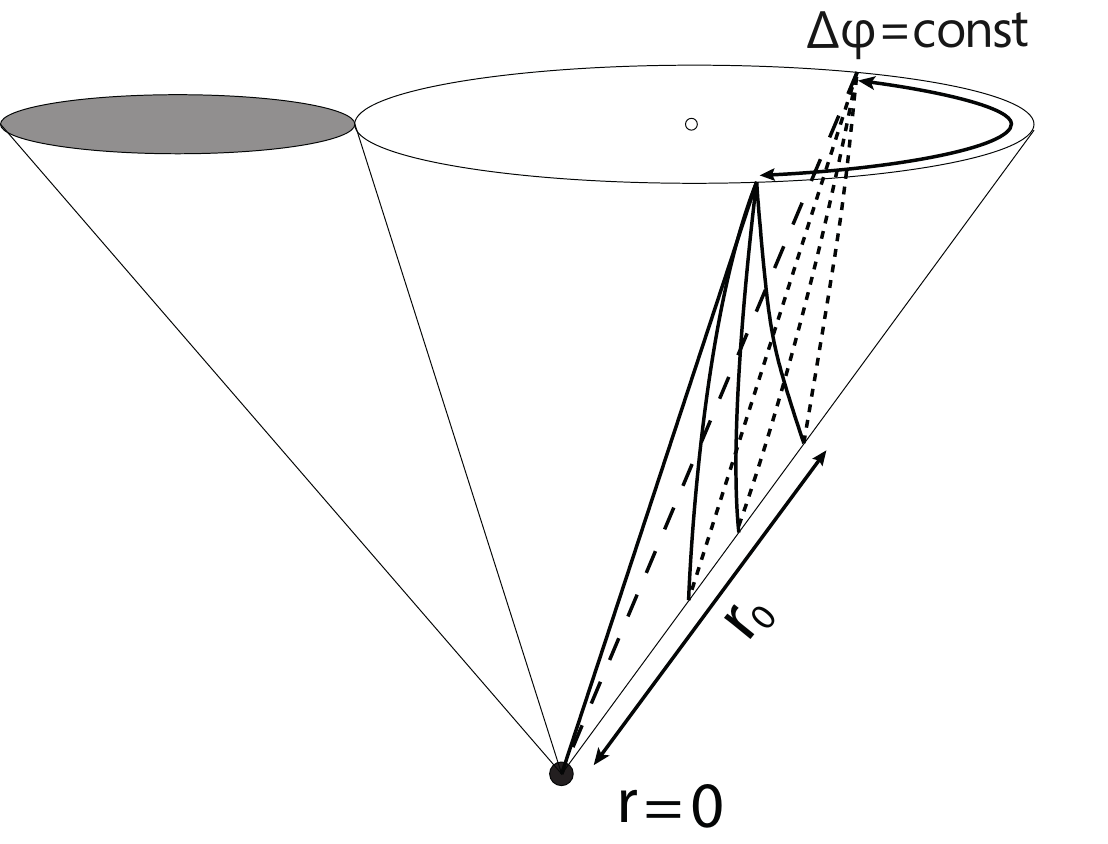}
        \includegraphics[width=.21\textwidth]{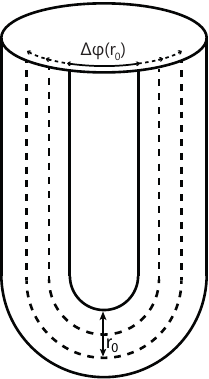}
    \caption{The asymptotic separation, $\Delta\varphi$, of the conifold D7 embedding (left) does not depend on the brane merging point, contrary to the Sakai-Sugimoto D8 embedding (right).}
    \label{fig:ComparingBraneEmbeddings}   
\end{figure}

A straightforward gravity calculation indeed leads to such a configuration. To be more precise, it produces a one-parameter family of D-brane profiles. This parameter $r_0$ can be seen as the lowest radial position of the D7 brane along $AdS_5$. For $r_0=0$ the profile looks like a disconnected $D7$-$\bar{D}7$ pair, while for $r_0>0$ the two branes merge together at $r=r_0$, see Figure \ref{fig:ComparingBraneEmbeddings}. The former and the latter profiles are called the V-shape and U-shape respectively. Importantly, for any value of the parameter $r_0$ the asymptotic UV separation between the brane and the anti-brane is $r_0$-independent. In other words, the parameter corresponds to a normalizable mode - a VEV in the dual gauge theory. The arbitrariness of the parameter corresponds to a flat direction which indicates the spontaneous breaking of the conformal symmetry. We stress again that these solutions exist in the probe limit and backreaction will probably spoil the fixed asymptotic separation.

For the V-shape the conformal symmetry remains unbroken as the induced metric on the D7 brane is just $AdS_5 \times S^3$. For the U-shape the conformal symmetry is broken spontaneously and the corresponding massless Goldstone boson indeed appears in the spectrum of the D7 brane fluctuations (as expected this mode is just the radial fluctuation around the merging point). Furthermore, the chiral (flavour) symmetry associated with the $D7$-$\bar{D}7$ pair is also broken spontaneously like in the Sakai-Sugimoto model, where a $D8-\bar{D8}$ pair similarly merges to form a single profile, see Figure \ref{fig:DpSeparation}. The important difference between the two models, though, is the asymptotic separation of the branes. This is related to the fact that the Sakai-Sugimoto brane setup is based on Witten's D4 background compactified on $S^1$, which has no conformal symmetry to begin with. Also contrary to the Klebanov-Witten background, supersymmetry is broken in Witten's model even without the embedding of the probe $D8$ branes. In the D7 setup, on the other hand, the presence of both $D7$ and $\bar{D7}$ breaks supersymmetry, as one can truly see from the non-holomorphicity of the embedding \footnote{This is true in general for $D7$ embeddings but not for other branes}.

The same construction obviously does not work for the $AdS_5 \times S^5$ background, since the 5-sphere has no non-trivial cycle and, as a result, the D7 will ``shrink" to a point on the $S^5$. This can be avoided only if the boundary conditions at infinity are properly fixed, which is certainly inconsistent with U-shape we are aiming at.
Similarly, with no 2-cycle around there is no need for tadpole cancellation by an anti D7 like in the conifold scenario. The topology, thus, plays an essential role in the construction.

The main objective of this paper is to understand the unique properties of the U-shape embedding, and to find other examples that posses the same properties.
Another interesting motivation is related to the Higgs like dilaton explored by \cite{Bellazzini:2012vz}. They examine the possibility that the recently discovered Higgs-like resonance actually corresponds to the (pseudo)dilaton. They have shown that, in principle, the dilaton can have the correct observed couplings, but this requires fine-tuning and strong dynamical assumptions that will produce a light dilaton. We do not search for a massive dilaton here, but it is naturally related to our work as we elaborate in Section \ref{sec:Summary}.

Besides finding other embeddings we were also interested in calculating the $a$-term anomaly coefficient from the dilaton action. As was shown in the recent proof of the $a$-theorem\cite{Komargodski:2011vj} the dilaton has information about the Weyl anomaly even in flat space. Thus if the massless mode we find in the spectrum is indeed the Goldstone mode of the conformal symmetry we expect to find the relevant coefficient. Although it is easy (as we review in Section \ref{sec:dilaton}) to calculate the $a$-term for gauge branes, the dilaton effective action coming from flavour brane fluctuations is a formidable task. As we mention in the relevant section, besides being hard to compute, it naively seems that the $a$-term will depend on $\lambda$ (the gauge theory coupling) which should not be the case\footnote{We thank O. Aharony for pointing it to us.}. This leads us to conjecture that this is an artefact of the probe approximation: in the full calculation (including backreaction) the $a$-term should disappear as $\mathcal{O}\left( N_\textrm{f}/N_\textrm{c} \right) $.

Let us briefly summarize our results: we have explored different flavour embeddings mainly in $T^{1,1}$. We will also discuss the ABJM model\cite{Aharony:2008ug}. In the embeddings we explore, which wrap a non-trivial cycle and hence require also an anti-brane, SUSY is broken explicitly, while "chiral" (flavour) and conformal symmetry are broken spontaneously. All of the new embeddings we explore are dCFT (unlike the D7 in KW which is a CFT) in which the operator insertion is localized in one or more spatial coordinates.

Our main claim is that starting with a non-contractible cycle (topologically non-trivial) in the compact space, a D-brane and anti D-brane wrapping the cycle naturally arrive at U- and V-shape profiles. A brane embedding wrapping a contractible cycle is dual (with zero flux on the brane) to a relevant deformation with the dual operator turned on (in all examples known to us).
We note that in general, having non-zero fluxes, orbifolds and other ingredients may alter our conclusions that wrapping a contractible cycle leads to a mass term and explicit conformal symmetry breaking. \cite{Aharony:1998xz} provides an example of a solution describing VEV (zero mass) deformation based on branes wrapping a contractible cycle.

The paper is organized as follows. In Section \ref{sec:D7s} we review the $D7-\bar{D}7$ embedding in $AdS_5\times T^{1,1}$ model, compare it to similar embeddings that are not dual to spontaneous conformal symmetry breaking, and discuss its unique features. In Section \ref{sec:Examples} we list possible candidates that potentially exhibit a U-shape embedding and show explicitly that indeed one can find such solutions. Section \ref{sec:dilaton} is dedicated to the dilaton, found in models exhibiting spontaneous conformal symmetry breaking. We also mention the unresolved issue with the dilaton $a$-term coefficient for probe flavour branes which we briefly discussed above.

\section{Mass and VEV deformation by probe D7-Branes}
\label{sec:D7s}

We start our journey with a very brief review of the D7 embedding in $AdS_5 \times S^5$ background originally studied in \cite{Karch:2002sh},\cite{Kruczenski:2003be}. The setup arises from the system of parallel and separated D3 and D7 branes. The separation parameter sets the mass of the quarks living on the 3-branes.
In the decoupling limit of the D3 branes, the asymptotic separation distance between the branes is related to the lowest position of the D7 along $AdS_5$.

The AdS and the 5-sphere metrics are:
\begin{equation}
\text{d} s_{AdS_5}^2 = r^2 \text{d} x_\mu \text{d} x^\mu + \dfrac{\text{d} r^2}{r^2} \,
\qquad
\text{d} s_{S^5}^2 = \text{d} \Phi^2 + \sin^2 \left( \Phi \right) \textrm{d} \alpha^2 +  \cos^2 \left( \Phi \right) \textrm{d} \Omega_{3}^2 \, .
\end{equation}
The angle $\Phi$ denotes the position of the 3-sphere inside the 5-sphere. In the induced metric on the world-volume of a D7 brane wrapping the 3-sphere,  the angle depends on the AdS coordinates $x_\mu$ and $r$. The DBI action reduces to the following action for  $\Phi(r,x_\mu)$\footnote{$\varphi^\star [g_{D7}]$ is  the pullback of the metric to the brane world-volume.}:
\begin{equation}
\label{L-Phi}
    \mathcal{L} = - \dfrac{\mu_{D7}}{{g_s}}\sqrt{-\det (\varphi^\star [g_{D7}])} \sim r^3 \cos^3(\Phi)\sqrt{1 + r^2 \dot{\Phi}^2 + \dfrac{1}{r^2} (\partial \Phi)^2  } \, ,
\end{equation}
where $\partial$ stands for the derivative with respect to the $4d$ coordinates and $\dot{\Phi}$ denotes the $r$-derivative of  $\Phi$. The EOM is solved by $r \cdot \sin (\Phi)=c$ \cite{Kruczenski:2003be}, so that the D7 extends in the IR up to $r_{\rm min}=c$. In the UV the asymptotic form of the solution is:
\begin{equation}
\Phi \sim c r^{-1} + \dfrac{c^3}{6} r^{-3} + \ldots \, ,
\end{equation}
Clearly, the solution describes a scalar dual to a dimension one operator. The fact that the non-normalizable $r^{-1}$ mode is switched on indicates that while solving the EOM we have to add a boundary term for the action (\ref{L-Phi}) at $r$-infinity. $r_{\rm min}=c$ is a free parameter and, as expected, corresponds to the D3-D7 separation distance, which is, in turn, equivalent to the ($\alpha^\prime$ rescaled) mass parameter.

\begin{figure}
    \centering
    \includegraphics[width=0.2\textwidth]{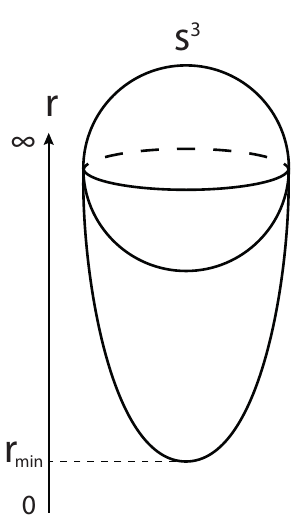}
    \caption{D7 flavour brane embedding on $S^5$ as it appears when flowing in the radial direction. In the IR the $S^3$ contracts inside the $S^5$ (it is a trivial cycle). The radial position where the brane terminates is dual to integrating out the massive quark.}
    \label{fig:D7inS5Profile}
\end{figure}

The D7-profile, therefore, does not have the U-form structure we described in Introduction. Its schematic behaviour is shown on Figure \ref{fig:ComparingBraneEmbeddings}. At the lowest point of the configuration $\Phi=\pi/2$, the 3-sphere shrinks and so the D7 brane tension goes to zero.

The asymptotic form of $\Phi(r)$ is still puzzling, however, due to the non-zero normalizable $r^{-3}$ term, dual to the VEV, which is not allowed on supersymmetry grounds (the brane setting preserves $\mathcal{N}=2$ SUSY). The puzzle was solved in \cite{Karch:2005ms}. They argued that for an interacting Lagrangian like \eqref{L-Phi}, the ``standard" UV expansion is not sufficient to determine the VEV. Instead one has to use the full ADS/CFT dictionary. Furthermore, the on-shell action is singular and has to be regularized by proper boundary counterterms that may further modify the VEV. For \eqref{L-Phi} there are only four counterterms that can have a non-zero contribution to the on-shell action: the $4d$ cosmological term $\sqrt{-\gamma}$, $\sqrt{-\gamma} {\gamma}^{\mu\nu} \partial_\mu \Phi \partial_\nu \Phi $,  $\sqrt{-\gamma} \Phi^2$ and $\sqrt{-\gamma} \Phi^4$, where $\gamma$ is the $4d$ induced metric $r^2 \text{d} x_\mu \text{d} x_\nu$.\footnote{Additional possible counterterms listed in \cite{Karch:2005ms} include $4d$ curvature which are strictly zero in this case.} The first three are necessary in order to cancel three independent divergences. The last one, on the other hand, produces only a finite contribution and its coefficient cannot be determined from regularity in the UV. In fact a finite term remains in the on-shell/energy calculation (with a contribution from the IR). The finite counterterm is then fixed to set the energy to zero, as should be if a supersymmetric preserving scheme is used. As a direct consequence of this scheme the VEV is also set to zero.

Let us now compare these results with the D7/anti-D7 model studied in \cite{Kuperstein:2008cq}. The motivation there was finding a U-shape solution similar to other known solutions in the literature, mainly Sakai-Sugimoto\cite{Sakai:2004cn}. The Sakai-Sugimoto model exhibits spontaneous chiral symmetry breaking, it is not conformal to begin with and also the UV separation depends on the merging point of the branes. A comparison between the conifold embedding which we review here and the Sakai-Sugimoto profiles is shown in Figure \ref{fig:ComparingBraneEmbeddings}.

The background solution was chosen to be the conifold (a cone over $T^{1,1}$), and as such the base's topology was $S^3\times S^2$. This was essential for the solution. Analogously to Sakai-Sugimoto the D7 branes (now wrapping the $S^3$), were essentially points in the transverse $S^2$. This embedding is not holomorphic (we will mention the holomorphic embedding below) and as such breaks all supersymmetry.

The absence of SUSY makes this a good candidate for displaying chiral symmetry breaking by a quark anti-quark condensate. This is because the QCD-like operator $\bar{\Psi}\Psi$ (we emphasize that this argument is relevant only for QCD-like quark anti-quark condensate), written as a part of a chiral superfield (where $Q=q+\theta \Psi+\dots$) is an F-term of $Q\tilde{Q}$, and as such if it acquires a VEV, SUSY must be broken - this is exactly the U-shape solution. There was in fact also a V-shape solution where the branes do not merge which also broke the supersymmetry completely. The fact that this model can be dual to a VEV deformation and is not restricted by SUSY makes this a good candidate also for conformal symmetry breaking. Of course a squark condensate could be formed which does not require SUSY to be broken but we suspect this is not the case here since this model and similar models we explore below break SUSY.

Now the orthogonal $5d$ metric is that of $T^{1,1}$:\footnote{In \cite{Kuperstein:2008cq} the metric was written using coordinates that make the $S^3 \times S^2$ structure explicit\cite{Gimon:2002nr}, but here for reasons of convenience we will stick to the more common form of the $T^{1,1}$ metric.}
\begin{equation}
\label{eq:KWMetric}
\text{d}s^2_{T^{1,1}}=\frac{1}{6}\sum_{i=1}^{2}\left(d\theta^2_i+\sin^2\theta_id\phi_i^2\right)+\frac{1}{9}\left(d\psi+\sum_{i=1}^{2}\cos\theta_id\phi_i\right)^2 \, .
\end{equation}
The topological 3-sphere is spanned, for example, by the angles $(\theta_1,\phi_1,\psi)$ and the 2-sphere by $\theta_2$ and $\phi_2$. The D7 embedding is then given by $\theta_2=\pi/2$ and $\phi_2=\phi(r)$. The pullback of the background metric to the probe brane gives the following DBI Lagrangian:
\begin{equation}
\label{eq:D7antiD7Lagrangian}
    \mathcal{L} \sim r^3 \left( 1 + \frac{r^2}{6} \phi^2_r \right)^{1/2} \, .
 \end{equation}
The EOM is solved by:
\begin{equation}
    \cos \left( \frac{4}{\sqrt{6}}\phi(r) \right) = \left(\frac{r_0}{r}\right)^4 \, .
\end{equation}
As advertised in the Introduction, the asymptotic UV separation is an $r_0$-independent constant $\Delta\phi=\frac{\sqrt{6}}{4}\pi$ (see Figure \ref{fig:ComparingBraneEmbeddings}). For $r_0=0$ one finds a V-shape of two separated branches, $D7$ and $\bar{D}7$.  For $r_0>0$ the two branches merge at $r=r_0$. The brane anti-brane interpretation naturally follows from the fact that the brane world-volume has opposite orientations as one approaches the two asymptotic points $\phi=\frac{\sqrt{6}}{8}\pi$ and $\phi=-\frac{\sqrt{6}}{8}\pi$. As a result, the net D7 charge on the 2-sphere is zero. The $\sqrt{6}$ factor here is a reminiscence of the fact that although the topology is that of $S^2\times S^3$, the 3-sphere is actually a squashed 3-sphere.

The induced metric on the D7 has an AdS factor only for $r_0=0$ and so $r_0$ parametrizes the breaking of scale invariance. From the asymptotic expansion of $\phi(r)$:
\begin{equation}
 \phi(r) \approx \pm \frac{\sqrt{6}}{8}\pi  \mp \dfrac{\sqrt{6}}{4} \left( \dfrac{r_0}{r}\right)^4 + \ldots
\end{equation}
we see that a $\Delta=4$ marginal operator acquires a VEV fixed by $r_0$:
\begin{equation}
\langle  O \rangle \sim \frac{r_0^4}{{\alpha^{\prime}}^2} \, .
\end{equation}
The fluctuation around this solution should give the corresponding Goldstone boson, the dilaton. The calculation was performed in \cite{Kuperstein:2008cq} and such a $4d$ massless mode was successfully identified (see \cite{Bayona:2010bg} and \cite{Ihl:2010zg} for a more detailed meson spectrum calculation).

In contrast to the previous case, the Lagrangian does not have $\cos(\phi)$ or any other term that introduce non-trivial interactions in the UV, and so we can conclude that the solution describes only the VEV deformation based exclusively on the asymptotic form of the solution. To be absolutely sure about this point, though, we have to address the issue of necessary (and all other possible) counterterms.

The Lagrangian (\ref{eq:D7antiD7Lagrangian}) diverges at infinity as $r^4$ and so we have to add counterterms. The situation is, however, conceptually different from the $AdS_5 \times S^5$ example, since the cosmological term alone can cancel the divergence, the $(\partial \phi)$ does not contribute, while all possible terms of the form $\phi^n$ may produce only finite contributions. Since these finite terms explicitly break conformal invariance we can omit them. To summarize, the brane action is regularized by a single counterterm. It is then straightforward to verify that solution still describes only a VEV deformation.

As a consistency check, we can verify that the on-shell action is indeed $r_0$-independent as it should be for a solution corresponding to a normalizable VEV mode. Adding the cosmological $r^4$-term (with a proper coefficient) we get:
\begin{equation}
\mathcal{L}_{\rm reg} \sim \int_{r_0}^{r_\Lambda}dr\frac{r^7}{(r^8-r_0^8)^{1/2}} -  \dfrac{r_{\Lambda}^4}{4} =  \frac{1}{4}(r^8-r_0^8)^{1/2}\bigg|^{r_\Lambda}_{r_0}   - \dfrac{r_\Lambda^4}{4}  = 0 \, ,
\end{equation}
where $r_\Lambda$ is the UV cut-off.

An alternative way to perform the check is to consider the difference between two on-shell actions for two different $r_0$'s. As one can easily check the difference also goes to zero as we take the cut-off to infinity.
In Sakai-Sugimoto this will not be the case as there is a non-normalizable solution as can be seen from dependency of the UV separation on the merging point (see \ref{fig:ComparingBraneEmbeddings}). This will also not be the case for $AdS_5 \times S^5$ since, as we reviewed above, the (unregulated) action depends on $r_\textrm{min}$.

This $D7$-$\bar{D}7$ setup is the only model known to us, which exhibits spontaneous conformal symmetry breaking by introducing a probe flavour D-brane (in fact, brane and anti-brane).

Let us also mention another D7 embedding on the conifold introduced in \cite{Ouyang:2003df} (see also \cite{Levi:2005hh}) . The embedding is holomorphic and so preserves all of the background supersymmetry. If we write the conifold equation as $z_1z_2-z_3z_4=0$, then
the embedding is given by $z_1=\mu$. Unlike the first example of this section, there is no brane construction here, and so there is no immediate way to identify $\mu$ as a mass parameter. Instead, it can be done either by identifying the holomorphic coordinates $z_i$ with the dual gauge theory fields or analysing the IR and the UV behaviour of the D7.  It appears that $\mu$ fixes not only the lowest position of the brane along the radial coordinate, but also also the non-normalizable mode in the UV. This means that the embedding does not look like a U-shape. This is consistent with the fact that the 3-cycle wrapped by the D7 is topologically trivial. A different holomorphic embedding was considered in \cite{Kuperstein:2004hy}, \cite{Cotrone:2010bv} with the profile equation $z_1 + z_2 =\mu$. For this setup $\mu$ is still a mass parameter and likewise the topology of the embedding
is trivial.

\section{More U-shape examples}
\label{sec:Examples}

Following the above arguments and examples, we now turn to searching for additional U-shape embeddings. We are interested in conformal $AdS_5 \times M^5$ backgrounds with the compact space $M^5$ having a topologically non-trivial (non-contractible) cycle, and no background RR source for the brane gauge fields. The last statement is relevant only for branes which couple linearly to the background flux, meaning the induced Wess-Zumino term has the form $\int P[C_{p}]\wedge F=\int P[F_{p+1}]\wedge A$.

\begin{table}
    \begin{center}
        \begin{tabular}{l*{14}{c}c}
             D$p$             & 0 & 1 & 2 & 3 & 4($r$) & 5 & 6 & 7 & 8 & 9 & $AdS_i \subset AdS_5$  & Cycle in $T^{1,1}$& FT & Type\\
            \hline
             D3                          & - & - & - & - & $\times$ & $\times$ & $\times$ & $\times$ & $\times$ & $\times$ & --- & --- & CFT &  gauge \\
             D3                           & - & $\times$ & $\times$ & $\times$ & - & - & - & $\times$ & $\times$ & $\times$ & 2 & 2 & dCFT & flavour \\
             D5                          & - & - & - & $\times$ & - & - & - & $\times$ & $\times$ & $\times$ & 4 & 2 & dCFT & flavour\\
             D5                           & - & - & $\times$ & $\times$ & - & - & - & - & $\times$ & $\times$ & 3 & 3 & dCFT & flavour\\
             D7                           & - & - & - & - & - & - & - & - & $\times$ & $\times$ & 5 & 3 & CFT & flavour\\
        \end{tabular}
        \caption{Topologically non-trivial embeddings in $AdS_5 \times T^{1,1}$ (KW).}
        \label{tab:KW}
    \end{center}
\end{table}

\begin{table}
    \begin{center}
        \begin{tabular}{c*{13}{c}c}
             D$p$              & 0 & 1 & 2 & 3($r$) & 4 & 5 & 6 & 7 & 8 & 9 & $AdS_i \subset AdS_4$  & Cycle in $\mathbb{CP}^{3}$ & FT & Type\\
            \hline
             D2                           & - & - & - & $\times$ & $\times$ & $\times$ & $\times$ & $\times$ & $\times$ & $\times$ & --- & --- & CFT & gauge\\
             D4                           & - & - & $\times$ & - & - & - & $\times$ & $\times$ & $\times$ & $\times$ & 3 & 2 & dCFT & flavour\\
             D6                           & - & - & $\times$ & - & - & - & - & - & $\times$ & $\times$ & 3 & 4 & dCFT & flavour\\
        \end{tabular}
        \caption{Topologically non-trivial embeddings in $AdS_4 \times \mathbb{CP}^{3}$ (ABJM).}
        \label{tab:ABJM}
    \end{center}
\end{table}

For the well-studied $AdS_5\times S^5$ geometry there is, of course, no such non-contractible cycle (a configuration that does not wrap at all the compact space just give the gauge brane configuration).
Motivated by the $D7$-$\bar{D}7$ model we will focus on the $AdS_5\times T^{1,1}$ background with type IIB branes in it. All of the examples in this category can be directly generalized to $Y^{p,q}$ 
\cite{Gauntlett:2004yd}, \cite{Martelli:2004wu}, \cite{Bertolini:2004xf}, \cite{Benvenuti:2004dy} and $L^{a,b,c}$ \cite{Cvetic:2005ft}, \cite{Benvenuti:2005ja}, \cite{Franco:2005sm}, \cite{Butti:2005sw}, \cite{Cvetic:2005vk} geometries that share the same $S^3 \times S^2$ topology as $T^{1,1}$.\footnote{See \cite{Canoura:2005uz} and \cite{Canoura:2006es} for supersymmetric probes in these geometries.} 

The type IIA examples will be based on the ABJM \cite{Aharony:2008ug} geometry, $AdS_4 \times \mathbb{CP}^{3}$. Although topologically $\mathbb{CP}^{3}$ is not a direct product of spaces like $T^{1,1}$, it has even dimensional cycles suitable for D4 and D6 branes.

In Tables \ref{tab:KW} and \ref{tab:ABJM} we have summarized all the possible candidates based on these two backgrounds. We indicated which AdS space the brane wraps, the dimensions of the compact cycles, what kind of field theory is expected (CFT or defect CFT) and whether the probe is a flavour brane or a gauge brane (the gauge branes are brought here for comparison). The co-ordinate denoted with ($r$) is the corresponding background radial coordinate.

We will present below all of the flavour brane embeddings in Table \ref{tab:KW} except the last one that has already been treated above. As for Table \ref{tab:ABJM} we will address only the D4 probe setup with the $\mathbb{CP}^{1} \in \mathbb{CP}^{3}$ cycle. The D6 case looks rather similar  and we will not report it here.

We believe that these tables exhaust all the possibilities for U-shape solutions in current known conformal backgrounds (as we have already pointed out, the $T^{1,1}$ construction can be generalized to $L^{a,b,c}$ geometries that have exactly the same $S^3 \times S^2$ topology). All of the new embedding are dual to defect CFTs. This is in contrast to the theory explored in \cite{Kuperstein:2008cq} where the brane wrapped the entire dual space-time coordinates. In the words of \cite{Aharony:2003qf}, the background fields (KW or ABJM) are the ambient fields and the probe branes are dual to the insertion of defect flavours to theory. The flavour fields are localized in lower spacetime dimensions, and so are their interactions with the background. If there is a suitable Lagrangian description to these theories it will be necessarily composed of two parts:
\begin{equation}
\mathcal{L}_{\hbox{dual dCFT}}=\mathcal{L}_{\hbox{background}}+\delta(x) \cdot \mathcal{L}_{\hbox{flavour and interactions}} \, ,
\end{equation}
where $\delta(x)$ represents the directions in which the defect flavours are localized. We have not been able to say much more on the dual theories than already analysed in \cite{Kuperstein:2008cq}. The dual theories should be similar, except of course that now we have a defect. As in the original model there is a quiver theory and spontaneous ``chiral" (flavour) symmetry breaking. Supersymmetry is again broken explicitly by the embeddings.

In the next subsections we show explicitly that U-shape solutions indeed exist for these embeddings.

\subsection{D5 probe wrapping $AdS_3\times S^3$ in $AdS_5\times T^{1,1}$}
\label{sec:ExplicitCalculations}

The simplest candidate in the  table above, is the D5 brane wrapping $AdS_3\times S^3$ in the $T^{1,1}$ background (No.4 in Table \ref{tab:KW}). Because the D5 brane wraps the same $S^3$ as in \cite{Kuperstein:2008cq}, the action will be almost the same. The Lagrangian density is now:
\begin{equation}
\label{eq:D5antiD5Lagrangian}
\mathcal{L} \sim r \left(1 + \frac{r^2}{6} \phi^2_r \right)^{1/2} \, .
\end{equation}
It essentially takes the same form as (\ref{eq:D7antiD7Lagrangian}) except for powers of $r$.
The solution will now be:
\begin{equation}
\phi(r) = \dfrac{\sqrt{6}}{2} \left( \pm \frac{\pi}{2}  \mp  \arctan \left( \dfrac{r_0^2}{\sqrt{r^4-r_0^4}} \right) \right) \, .
\end{equation}
This obviously satisfies the same boundary conditions as before, but now the brane UV separation is $\frac{\sqrt{6}}{2} \pi$ rather than $\frac{\pi \sqrt{6}}{4}$ of \cite{Kuperstein:2008cq} . So we have found another U-shape solution, but this time the dual theory must be a defect CFT. The ambient theory is the  four dimensional field theory  of \cite{Klebanov:1998hh} with additional flavour quark fields that reside only in two out of the four space-time dimensions, namely along the time and the space directions which are wrapped by the flavour probe brane. This near boundary behaviour is that of a VEV of a marginal (meaning $\Delta=2$) operator in 2 space-time dimensions, as expected.

\subsection{D5 probe wrapping $AdS_4\times S^2$ in  $AdS_5\times T^{1,1}$ }
\label{sec:D5AdS4S2}

This solution should have the usual near boundary behaviour of a free scalar in $AdS_4$ dual to a dimension 3 (marginal) operator. This brane embedding was found very recently  by Filev, Ihl and Zoakos in \cite{Filev:2013vka}. Their solution is, in turn, based on the supersymmetric D5 embeddings found earlier in \cite{Arean:2004mm} as we will review in a moment.

In terms of the coordinates used in the metric (\ref{eq:KWMetric}) the embedding of \cite{Filev:2013vka} is defined by:\footnote{There is an additional option with $\theta_2=\theta_1 \, , \,\,\phi_2= 2\pi-\phi_1$.}
\begin{equation}
\label{eq:D5-Ansatz}
\theta_2 = \pi - \theta_1 \,, \quad  \phi_2 = \phi_1 \, ,
\quad \textrm{and} \quad \psi=\psi(r) \, .
\end{equation}
Notice that with this Ansatz the induced metric looks simpler, as the fibre mixed terms disappear. Consequently one can assume that $\psi$ depends only on $r$. The Lagrangian density is:
\begin{equation}
\mathcal{L} \sim r^2 \left(1+\frac{1}{9}r^2\dot{\psi}^2 \right)^{1/2} \, .
\end{equation}
The EOM is then:
\begin{equation}
\dfrac{r^4\dot{\psi}}{\sqrt{1+\dfrac{r^2\dot{\psi}^2}{9}}} = \textrm{const}
\end{equation}
and the solution is:
\begin{equation}
\psi(r)=\pm \frac{\pi}{2}  \mp \arcsin\left(\frac{r_0^3}{r^3}\right) \, .
\end{equation}
Again, the UV behaviour is the one we expect for a $\Delta=3$ VEV deformation.

Let us now complete this example by demonstrating that this ansatz indeed corresponds to D5 wrapping the 2-sphere. The conifold is defined by a complex $2 \times 2$ matrix $W$ with $\text{det} W=0$. The radial coordinates on the conifold is determined by $2 r^2 = \textrm{Tr} \left( W^\dagger W \right)$. Any such
degenerate matrix can be written as $W =\sqrt{2} r u v^\dagger$, where $u$ and $v$ are complex unit vectors ($u^\dagger u = v^\dagger v = 1$) quotiented by $(u,v) \to e^{i \varphi} (u,v)$. Using $u$ and $v$ one can define\footnote{See \cite{Evslin:2007ux}, \cite{Krishnan:2008gx} and \cite{Kuperstein:2008cq} for more details.} a matrix $X \in SU(2)$ by $u = X v$, which has a unique solution\footnote{Here $\epsilon = \left( \begin{array}{cc} 0 & 1 \\ -1 & 0 \end{array} \right)$.}
$X = u v^\dagger - \epsilon u v^\text{T} \epsilon$. Since $X$ is invariant under the $\varphi$-quotient it defines an $S^3$. On the other hand, $v$ is still defined modulo $v = e^{i \varphi} v$ and so describes an $S^2$. Starting with $r$, $X$ and $v$ we can fix $u$ and therefore $W$, and vice versa. In terms of the angles in (\ref{eq:KWMetric}) we have:
\begin{equation}
u = e^{\frac{i}{4} \psi} \left( \begin{array}{c} \cos \frac{\theta_1}{2}  \, e^{\frac{i}{2} \phi_1} \\ \sin \frac{\theta_1}{2}  \, e^{- \frac{i}{2} \phi_1}  \end{array} \right)
\quad
\text{and}
\quad
v = e^{- \frac{i}{4} \psi} \left( \begin{array}{c} -\sin \frac{\theta_2}{2}  \, e^{-\frac{i}{2} \phi_2} \\ \cos \frac{\theta_2}{2}  \, e^{\frac{i}{2} \phi_2}  \end{array} \right)
\, .
\end{equation}
The Ansatz (\ref{eq:D5-Ansatz}) then implies that for $\psi = \pm \dfrac{\pi}{2}$:
\begin{equation}
X = \pm i \left( \begin{array}{cc} -1 & 0 \\ 0 & 1 \end{array} \right) \, .
\end{equation}
In other words, for the two asymptotic values of $\psi$ we find two different points on the 3-sphere parametrized by $X$. In fact, the two solutions with $\psi=\frac{\pi}{2}$ and $\psi=-\frac{\pi}{2}$ were found already in \cite{Arean:2004mm}. They both correspond to supersymmetric D5 embeddings but they preserve different supersymmetries, confirming our $D5$-$\bar{D}5$ interpretation. Away from the UV, for $\psi \neq \pm \frac{\pi}{2}$, it seems that $X$ will not be constant for a fixed $r$ and will depend on the $S^2$ angles $\theta_2$ and $\phi_2$. Notice, however, that we can properly adjust the definition of $X$, setting instead $u=A_0 X v$, where $A_0$ is a constant $U(2)$ matrix. With a proper choice of $A_0$ one can guarantee that $X$ is constant for a given value of $\psi(r)$. 
Unfortunately, we were not able to find an $S^3 \times S^2$ trivialization that works for any value of $\psi(r)$, similar to what one can do for the D7 brane embedding (see \cite{Kuperstein:2008cq}). Nevertheless, it is clear that the D5 brane indeed wraps a 2-cycle topologically equivalent to the $S^2$.

\subsection{D3 probe wrapping $AdS_2\times S^2$ in $AdS_5\times T^{1,1}$ }

Similarly to the D5 embedding above, we can now embed a D3 brane which wraps the same 2-cycle (No.2 in table \ref{tab:KW}). The derivation of the Lagrangian density is the same as for the similar D5 embedding so we will not repeat it. The only difference now is the radial factor from the $AdS_2$, meaning the EOM is:
\begin{equation}
\dfrac{r^2\dot{\psi}}{\sqrt{1+\dfrac{r^2\dot{\psi}^2}{9}}} = \text{const} \, .
\end{equation}
Using the same boundary conditions as before we get:
\begin{equation}
\psi(r)=\mp\frac{3\pi}{2}  \pm 3\arctan\left(\frac{r_0}{\sqrt{r^2-r_0^2}}\right) \, ,
\end{equation}
which is the expected near boundary behaviour of a VEV deformation of a marginal ($\Delta=1$) operator in one dimension. Notice that here $\Delta\psi$ at the boundary is $3\pi$, which does not mean the brane wraps the cycle more the once, since $\psi\in[0,4\pi)$.
In the defect dual field theory the additional flavoured quark fields are ``quantum mechanical" in the sense that they are only time dependent but space independent.

This is different from the usual $D3$ gauge branes. The branes now reach the boundary, making them flavour branes as we mentioned in the Introduction.

\subsection{D4 probe wrapping $AdS_3 \times \mathbb{CP}^{1}$ in $AdS_4 \times \mathbb{CP}^{3}$}

The complex projective space $\mathbb{CP}^{3}$ is given by vectors in $\mathbb{C}^{4}$ identified up to an overall complex rescaling. We parametrize the homogeneous $\mathbb{C}^{4}$ coordinates as in \cite{Murugan:2011zd}:
\begin{eqnarray} \label{homogenous-cp3}
\nonumber && z^{1} = \cos{\zeta}\sin{\tfrac{\theta_{1}}{2}} ~ e^{i\left(y + \frac{1}{4}\psi - \frac{1}{2}\phi_{1}\right)} \hspace{1.0cm}
z^{2} = \cos{\zeta}\cos{\tfrac{\theta_{1}}{2}} ~ e^{i\left(y + \frac{1}{4}\psi + \frac{1}{2}\phi_{1}\right)} \\
&& z^{3} = \sin{\zeta}\sin{\tfrac{\theta_{2}}{2}} ~ e^{i\left(y - \frac{1}{4}\psi + \frac{1}{2}\phi_{2}\right)} \hspace{1.05cm}
z^{4} = \sin{\zeta}\cos{\tfrac{\theta_{2}}{2}} ~ e^{i\left(y - \frac{1}{4}\psi - \frac{1}{2}\phi_{2}\right)},
\end{eqnarray}
with $\zeta \in [0,\tfrac{\pi}{2}]$, $\theta_{i} \in [0,\pi]$, $\psi \in [0,4\pi]$ and $\phi_{i} \in [0,2\pi]$.  Moreover, $\sum_{A} \left\vert z_z \right\vert^2=1$ and $y \in [0,2\pi]$ is the common phase on which the inhomogeneous  $\mathbb{CP}^{3}$  coordinates ${z^{i}}/{z^{4}}$ do not depend.

In these coordinates $\zeta$, $\theta_{i}$, $\psi$ and $\phi_{i}$, the Fubini-Study metric of $\mathbb{CP}^{3}$ is given by
\begin{eqnarray}
\nonumber & \text{d}s_{\mathbb{CP}^{3}}^{2} = & d\zeta^{2}
+ \frac{1}{4}\hspace{0.05cm}\cos^{2}{\zeta}\sin^{2}{\zeta}
\left[d\psi + \cos{\theta_{1}}d\phi_{1} + \cos{\theta_{2}}d\phi_{2}\right]^{2} \\
&& + \frac{1}{4}\hspace{0.05cm}\cos^{2}{\zeta}\left(d\theta_{1}^{2} + \sin^{2}{\theta_{1}}d\phi_{1}^{2}\right)
+ \frac{1}{4}\hspace{0.05cm}\sin^{2}{\zeta}\left(d\theta_{2}^{2} + \sin^{2}{\theta_{2}}d\phi_{2}^{2}\right) \, .
\end{eqnarray}
Type IIA string theory on the background $AdS_{4}\times \mathbb{CP}^{3}$ has the following metric:
\begin{equation}
\text{d}s^{2} = R^2\left(\text{d}s_{AdS_{4}}^{2} + 4 \text{d}s_{\mathbb{CP}^{3}}^{2}\right) \, .
\end{equation}
Adopting an ansatz similar to the D5 embedding above, we take the brane to wrap $\{x_0,x_1,r,\theta=\theta_1,\phi=\phi_1\}$ and the transverse embedding to be $\{\theta_2=\theta,\phi_2=2\pi-\phi,\psi=0,x_2=0,\zeta(r)\}$.
The induced metric is then:
\begin{equation}
\text{d}s_{D4}^{2} =r^2 \text{d} x_{\mu} \text{d} x^{\mu}+ \text{d} r^2\left(\frac{1}{r^2}+4\dot{\zeta}^{2}(r)\right) +\left(\text{d}\theta^{2} + \sin^{2}{\theta}\text{d}\phi^{2}\right) \, .
\end{equation}
The Lagrangian density is then:
\begin{equation}
\mathcal{L} \sim r\left(1+4r^2\dot{\zeta}^2(r)\right)^{1/2} \, .
\end{equation}
The EOM is then:
\begin{equation}
\frac{4r^3\dot{\zeta}}{\sqrt{1+4r^2\dot{\zeta}^2}}=\text{const} \, .
\end{equation}
Using the same boundary conditions as before we get:
\begin{equation}
\zeta(r) = \pm \frac{\pi}{8} \mp \frac{1}{4} \arctan\left(\frac{r_0^2}{\sqrt{r^4-r_0^4}}\right)
\end{equation}
and at the boundary $\Delta\zeta=\frac{\pi}{4}$.

To see that the brane really wraps $\mathbb{CP}^{1}$, we notice that for this embedding:
\begin{eqnarray}
\nonumber && z^{1} = \cos{\zeta}\sin{\tfrac{\theta}{2}} ~ e^{i\left(y + \frac{1}{2}\phi\right)} \hspace{1.0cm}
z^{2} = \cos{\zeta}\cos{\tfrac{\theta}{2}} ~ e^{i\left(y - \frac{1}{2}\phi\right)} \\
&& z^{3} = \sin{\zeta}\sin{\tfrac{\theta}{2}} ~ e^{i\left(y  + \frac{1}{2}\phi\right)} \hspace{1.05cm}
z^{4} = \sin{\zeta}\cos{\tfrac{\theta}{2}} ~ e^{i\left(y - \frac{1}{2}\phi\right)}
\end{eqnarray}
These are obviously not independent coordinates, and so we can use a new set of coordinates:
\begin{equation}
w_1=z_1+iz_3=\sin\frac{\theta}{2}~e^{i\left(y+\frac{1}{2}\phi+\zeta\right)} \, ,
\quad
w_2=z_2+iz_4=\cos\frac{\theta}{2}~e^{i\left(y-\frac{1}{2}\phi+\zeta\right)} \, .
\end{equation}
Clearly, $|w_1|^2+|w_2|^2=1$ and, together with the $U(1)$ quotient by $\zeta+y$, it is equivalent to $\mathbb{CP}^1=S^2$.

\section{The dilaton}
\label{sec:dilaton}

From the usual AdS/CFT dictionary our results suggest (perhaps only in the probe approximation) that conformal symmetry is broken spontaneously. If this is indeed the case the configuration should display the corresponding Goldstone boson - the dilaton. Other than identifying a massless mode in the spectrum, for even dimensions we can also use the restrictions on the effective action of the dilaton to identify the $a$-term coefficient\cite{Komargodski:2011vj}. This is of course relevant only for the space-time filling $D7$ brane and not to our new dCFT embeddings. We mention it here mainly because of the unresolved issue that is explained below.

It was shown in \cite{Komargodski:2011vj} that the effective action carries the anomaly term $S_{\rm anomaly}|_{g_{\mu\nu}=\eta_{\mu\nu}}\sim2a\int d^4x(\partial\tau)^4$. The UV theory we will have $a_{\rm UV}$ but flowing to the IR, by virtue of the anomaly matching, the coefficient will be $a_{\rm UV}-a_{\rm IR}$\footnote{This is true up to additional $a_{scalar}$ term in the IR from the dilaton}. \cite{Komargodski:2011vj} went further on to show that $a_{\rm UV}>a_{\rm IR}$, but here we are interested only in the coefficient itself. The dilaton effective action was calculated for various probe gauge brane embeddings in \cite{Elvang:2012st}. Their D3-brane analysis in $4d$ (dual to the $N=4$ Coulomb branch) can be repeated for any five dimensional Sasaki-Einstein space.

The $AdS_5$ metric in Poincar\'e coordinates is given by:
\begin{equation}
\text{d}s^2=\frac{R^2}{z^2} \left( \text{d}x^\mu \text{d}x_\mu + \text{d}z^2 \right) \, ,
\end{equation}
where $R^4=\frac{(2\pi)^4 g_s l_s^4}{4 \text{Vol}(M)}$. We are only interested in the dilaton, so we ignore motion of the brane in any of the internal coordinates, so the DBI induced action for the spacetime filling D3 brane is:
\begin{equation}
\label{eq:D3BraneDBIAction}
S_{D3}=-\mu_3\int{d^4x\frac{R^4}{z^4}\left(\sqrt{1+(\partial z)^2}-1\right)}\simeq\mu_3R^4 \int d^4x \left( -\frac{ ( \partial z )^2}{2z^4}+\frac{(\partial z)^4}{8z^4}+\cdots \right) \, .
\end{equation}
Here $\mu_3=\left((2\pi)^3g_s l_s^4 \right)^{-1}$ and the $(-1)$ following the square root comes from the 5-form flux. This way the no-force condition ensures the dilaton's flat potential. The last step required is the field redefinition $z=R \left( 1-\frac{\phi}{f} \right)^{-1}$ relating field $z$ to the canonically normalized dilaton $\phi$ and the decay constant $f$. All in all, one gets the following effective action:
\begin{equation}
S_{D3}=\int{d^4x \left( -\frac{\mu_3R^2}{2f^2}(\partial\phi)^2+\frac{\mu_3 R^4}{8f^4}(\partial\phi)^4+ \cdots \right) } \, .
\label{eq:PhiEffectiveAction} \, .
\end{equation}
According to \cite{Komargodski:2011vj}, $\Delta a=a_{\rm UV}-a_{\rm IR}=\frac{1}{16}\mu_3 R^4 = \frac{\pi N}{32\text{Vol}(M)}$. We see that \mbox{$\Delta a \sim \left(\text{Vol}(M)\right)^{-1}$},  as expected from \cite{Gubser:1998vd} .

For flavour branes, however, things are more complicated. As we have already seen, contrary to the gauge branes case, the flavour branes have an $r$-dependent profile making it harder to identify the four dimensional physical dilaton effective action. For example, the $D7$-$\bar{D }7$ embedding we discussed extensively above, gives the following five dimensional effective action (up to four derivatives):
\begin{multline}
    S=-\frac{\mu_72\pi^2}{72}\int d^4x d\mathfrak{z} \left(\frac{1}{2}(\partial_\mathfrak{z}\delta y)^2+\frac{R^4}{32r^{10}}\partial_\mu\delta y\partial^\mu \delta y\right.\\\left.-\frac{R^8}{8 \cdot 2^8r^{20}}(\partial_\mu \delta y)^4)+ \frac{1}{4}\frac{R^4}{32r^{10}}(\partial_\mu\delta y\partial^\mu \delta y)(\partial_\mathfrak{z}\delta y)^2-\frac{1}{8}(\partial_\mathfrak{z} \delta y)^4+\cdots\right) \, .
    \label{eq:KWExpansionOnlyDerivative}
\end{multline}
Here $r^8=y^2+\mathfrak{z}^2$, $\mu_7 = \left( (2\pi)^7 g_s l_s^8 \right)^{-1}$, and $R^4=\frac{27}{4}\pi g_s N_\text{c} l_s^4$.
This is not the four dimensional effective action yet, but we can already see an inherent issue here that we did not succeed to resolve. It seems that the $a$-term coefficient will always carry some extra factors of the string/gauge coupling coming from $\mu_7R^8\sim g_sN_c^2\sim \lambda N_c$. This issue is common to all dilatonic modes which may arise from flavour branes fluctuations, because of the mismatch between the brane tension and volume: both of these carry different powers of $g_s$: $\mu_7 \sim \left( g_s l_s^8 \right)^{-1}$ and $\text{Vol}_\text{D7}\sim R^8\sim g_s^2l_s^8$. This is in contrast to the $D3$ gauge brane where the tension and the volume have the same power of $g_s$. The $\Delta a$-term should not depend on the gauge coupling at the end of the flow. On the gravity side we have only the flavour brane tension $\mu_\textrm{flavour}$ and the AdS radius $R$, and building a dimensionless coefficient from these two will always contain some factor of $g_s$ or $l_s$ (or $\lambda$ when written in FT terms). Hence, it does not seem plausible that a simple field redefinition would solve this contradiction. 

We are led to suspect that in our approximation the $a$-term should be zero when the full 4-dim effective action is calculated, and that the affect of the flavour brane on the $a$-term will be visible only when backreaction is taken into account.


\section{Models with a pseudo-dilaton}
\label{sec:pseudodilaton}

\begin{figure}[htb]
	\begin{center} $
		\begin{array}{cc}
            \includegraphics[width=6.7cm]{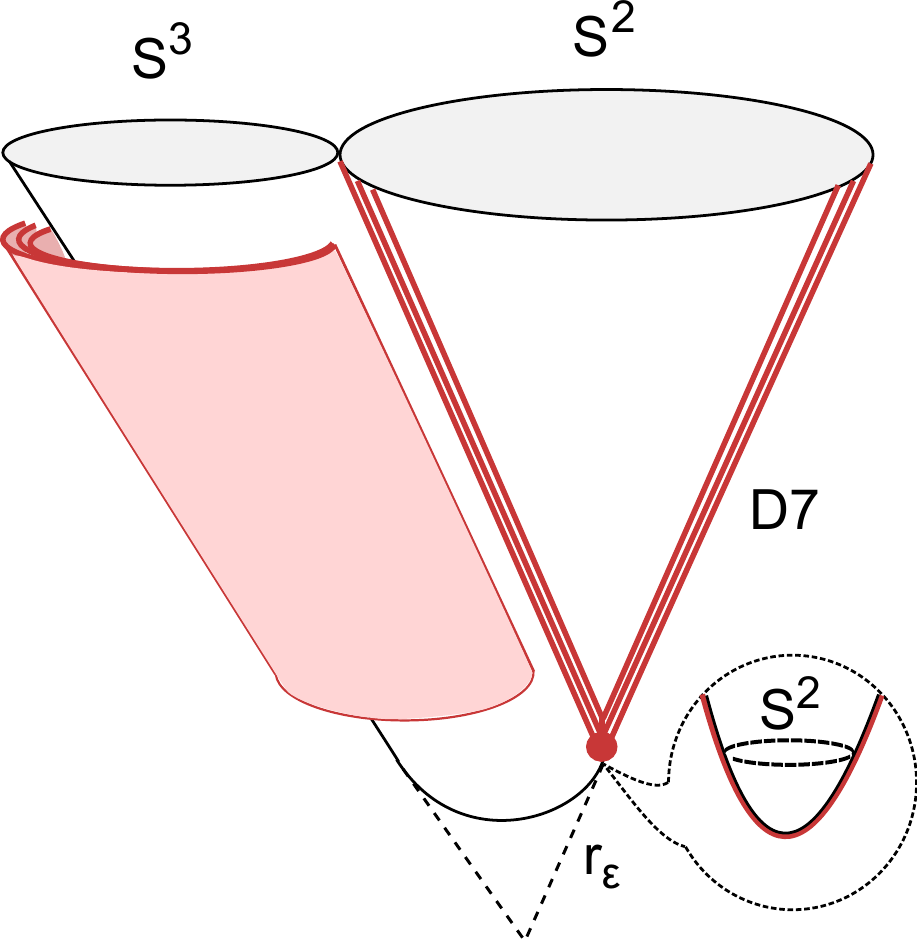}
		\quad & \quad
            \includegraphics[width=6.5cm]{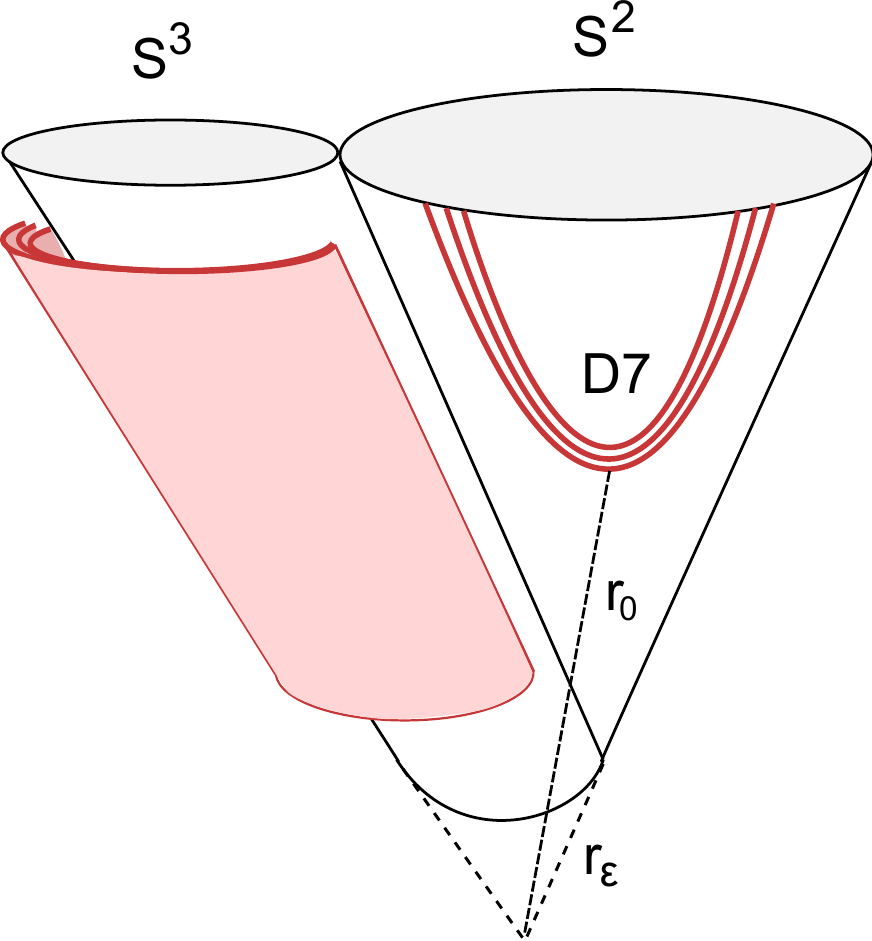}
		\end{array}$
	\end{center}
	\caption{Embedding of the D7-branes in the KS background. The branes wrap the $S^3$ and span a line in $R^1\times S^2$. The lowest point of the U-shape profile is $r_0$. The deformation scale is $r_\epsilon$. The figures illustrate the anti-podal (left) and the non antipodal (right) cases respectively \cite{Dymarsky:2010ci}.}
    \label{d7ks}
\end{figure}

Relating our holographic dilaton to the phenomenology of elementary particles  may be possible if one renders the dilaton into a massive pseudo dilaton.
Indeed, \cite{Bellazzini:2012vz} discussed the possibility that the recently discovered 125 GeV Higgs like resonance corresponds actually to a pseudo-dilaton. Since the dilaton couples to the trace of the energy-momentum tensor its couplings and hence decay modes are similar to those of the Higgs particle.
An important issue raised in this paper is how to get a pseudo-dilaton  lighter than the scale of the corresponding strong interaction.
This question was explored in \cite{Dymarsky:2010ci}\footnote{In fact the purpose  of having a lighter scalar meson than the corresponding vector mesons, was very different in that paper. The issue was how to get at large separation distances  an attractive force  of the  ``nuclear interaction" which is stronger than the repulsive one. The pseudo Goldstone mechanism enabled such a scenario.}. A pseudo-dilaton  corresponds to a system where on top of the spontaneous breaking there is also an explicit breaking of conformal invariance. In holography this can be achieved by     placing  a U-shape probe flavour branes in a background  for which the conformal symmetry is broken explicitly. Introducing a scale to  the KW background has been achieved in the deformed conifold solution of  Klebanov-Strassler \cite{Klebanov:2000hb}. Incorporating D7 anti-D7  branes in this background is technically much more complicated than in the KW background. This problem was first addressed in \cite{Sakai:2003wu} and finally resolved in \cite{Dymarsky:2009cm}.
The description of the D7 branes in this background is given in Figure \ref{d7ks}.

 The main difference between the model based on the KS background and that based on the KW one is that in KS case  there is an internal scale $r_\epsilon\equiv e^{2/3}$ (the size of the blown up $S^3$ at the tip) that breaks conformal symmetry explicitly. However, because there are two scales: $r_0$ which breaks the symmetry spontaneously and $r_\epsilon$ which breaks it explicitly, we expect a parametrically light pseudo-dilaton in the spectrum. Indeed, the spectrum of gauge singlets of this model include  glueballs with mass:
  \begin{equation}
  m_\textrm{glueball} \sim \frac{r_\epsilon}{\lambda \alpha^\prime} \, .
  \end{equation}
  At the same time,  for $r_0\gg r_\epsilon$ we can treat the effect of $r_\epsilon$ as a small correction, thus the masses of regular mesons remain essentially the same:
  \begin{equation}
  m_\textrm{meson} \sim \frac{r_0}{\lambda \alpha^\prime} \sim \frac{r_0}{r_\epsilon} m_\textrm{glueball} \, ,
  \end{equation}
  while the pseudo-dilaton acquires a small mass of the form:
  \begin{equation}
  m_\textrm{dilaton}  \sim \left(\frac{r_\epsilon}{r_0} \right)^2 m_\textrm{glueball} \, .
  \end{equation}
Obviously the dilaton becomes massless when we turn off the explicit breaking as expected.
The holographic model indeed obeys the phenomenological requirement  that the pseudo-dilaton is lighter than the dynamical scale of the system. It should be emphasized that in attempting to relate the holographic model to elementary particle phenomenology, the underlying large $N$ gauge symmetry does not relate to QCD, but rather to an additional strong interaction with a scale which  is much higher than that of QCD.

Now that we have clarified the potential relation of our prototype model with phenomenology, an obvious question is what is the situation  for the other holographic models discussed above.
The holographic duals of of the defect field theories listed in Table \ref{tab:KW} of Section \ref{sec:Examples}, are based on backgrounds with the transverse space which is the conifold. These cases could be transformed to the deformed conifold background of the KS background in exactly the same way as the prototype model of the D7 branes. As we mentioned there are also possible setups based on the Sasaki- Einstein manifolds $Y^{p,q}$ and $L^{a,a,c}$. It was shown in \cite{Franco:2005zu}, \cite{Berenstein:2005xa}, \cite{Bertolini:2005di} that supersymmetric system on such backgrounds cannot be deformed like in the conifold geometry. However, in this work we consider only non-supersymmetric models and hence the corresponding no-go theorem does not apply.

\section{Summary and open questions}
\label{sec:Summary}

 The main focus of this paper has been spontaneous conformal symmetry breaking due to flavour branes and anti-branes.
 Generalizing the spacetime filling $D7$ embedding of \cite{Kuperstein:2008cq}, we found (and reviewed) a number of  holographic models dual to defect field theories.
 
 The proposed construction relies on the conformal symmetry of the AdS part of the background and a non-trivial topology of the compact orthogonal space. The latter is necessary for the U-shape D-brane profile. For this reason the KW geometry, $AdS_5 \times T^{1,1}$, provides a fruitful playground for this approach in type IIB supergravity. All of the presented conifold based models can be straightforwardly generalized to $Y^{p,q}$ and $L^{a,b,c}$ spaces, since they also possess $S^3 \times S^2$ topology. For the type IIA model building we used $AdS_4 \times \mathbb{CP}^3$ background of the ABJM model, with D-branes wrapping the even dimensional cycles of $\mathbb{CP}^3$.
 
 Based on our findings we conjecture that for any compact space meeting the criteria of Section \ref{sec:Examples} there will be a U-shape probe brane solution that exhibits the features of spontaneous conformal symmetry breaking.

 The results of this paper are only the first step in the quest for a reliable holographic model. There are a handful of open questions that deserve further investigation. Let us list some of them.

 \begin{itemize}
 \item
 The most important open question is the issue of backreaction. The question is whether the Goldstone boson, the dilaton, will survive (remain massless) once backreaction is incorporated. In geometrical terms, it translates into the (in)dependence of the asymptotic separation between the two ends of the U-shape on the $r_0$ parameter, the tip location of the brane configuration. Differently stated, the massless Goldstone mode associated with the tip moving may acquire a mass beyond the probe approximation. 
 It was shown in \cite{Burrington:2007qd} that in the context of the Sakai-Sugimoto model the leading correction in $N_\text{f}/N_\text{c}$ does not change the structure of the U-shape configuration. However, since the results of \cite{Burrington:2007qd} deal with a different model and are limited only to the leading order correction, they cannot ensure the robustness of our mechanism. The backreaction might even spoil the solution altogether (for instance, it may lead to a runaway behaviour). We also mention that to our knowledge there are no known non-SUSY CFTs with a moduli space at finite $N$. The first steps in this direction were done in \cite{Ihl:2012bm} and \cite{Alam:2013cia}, where the first order backreaction was investigated in the smearing approximation.  
 \item
 Another issue, possibly related to the D-brane backreaction, is the coefficient of the  ``$a$-term"  in the dilaton effective action. As was explained in Section \ref{sec:dilaton}, the ``$a$-term"  is associated with the action of the mode suspected to be the dilaton, and its coefficient does not meet the ``$a$-theorem" expectations. It was found to be linear in $\lambda$, whereas according to the theorem it should be independent of the gauge coupling constant. Resolving this contradiction (for example, showing explicitly that the term vanishes identically) is an important open question left for future investigation.
 \item
 We still have to establish a map between the holographic setup and the dual field theory. In \cite{Kuperstein:2008cq} the quiver field theory associated with the prototype model was discussed. Further arguments from the field theory side to support the main idea will be  definitely welcomed.
 The holographic model admits two types of Goldstone bosons. The one discussed in this paper is associates with the fluctuation of the embedding and throughout the paper has been referred to as the dilaton.
 The second type  are the  massless modes of the flavour gauge fields that reside on the probe flavour branes. These modes are  the dual  of the Goldstone bosons associated with the spontaneous breaking of chiral symmetry. They were identified originally in the Sakai-Sugimoto model \cite{Sakai:2004cn} and later in \cite{Kuperstein:2008cq} for our KW  prototype model.  Thus the field theory dual of our models should contain both pions and a dilaton.
 \item
 One obvious direction of  further investigation is the correlation functions computation. The two point function of the dilaton has a very special signature. It has to behave like $1/p^2$. This type of behaviour was indeed observed in \cite{Hoyos:2013gma}. Finding these correlators in the setups of this paper will provide further important evidence for our basic idea. Such correlators were recently studied in \cite{Argurio:2013uba}.
 \item
 In Section \ref{sec:pseudodilaton} the pseudo-dilaton arising in the context of the Klebanov-Strassler model was  briefly discussed. It was argued in \cite{Dymarsky:2010ci} that one can have a pseudo-dilaton which is parametrically lighter than the scale of the strong interaction.
 As was mentioned in Introduction this was the main phenomenological motivation of our work. Obviously this paper just raises the possibility and a further deeper investigation is needed. In particular, it is important to analyse the couplings of the pseudo-dilaton mode to other modes and its decay channels.
 \item
 One may investigate the perturbative stability of the new D-brane profiles using the method proposed recently in \cite{Anguelova:2013lna}.
\end{itemize}

\acknowledgments{We would like to thank  Ofer Aharony for his insightful comments on the manuscript and  for useful discussions. We are also grateful to Veselin Filev and Matthias Ihl for the e-mail exchange about the results of their upcoming paper.
 The work of O.B and J.S. is partially supported by the Israel Science Foundation (grant 1665/10). S.K. is supported in part by the ANR grant 08-JCJC-0001-0 and the ERC Starting Grants 240210 - String-QCD-BH and 259133 -- ObservableString. }

\bibliographystyle{utphys}

\bibliography{final}

\end{document}